# Enhanced Photovoltaic Performances of Graphene/Si Solar Cells by Insertion of an MoS$_2$ Thin Film

*Yuka Tsuboi[†], Feijiu Wang[†], Daichi Kozawa[†], Kazuma Funahashi[‡],*

*Shinichiro Mouri[†], Yuhei Miyauchi[†§], Taishi Takenobu[‡] and Kazunari Matsuda[†]\**

[†]Institute of Advanced Energy, Kyoto University, Gokasho, Uji, Kyoto 611-0011, Japan

[‡]Department of Advanced Science and Engineering, Waseda University, Shinjuku-ku, Tokyo 169-8555, Japan

[§]JST, ERATO, Itami Molecular Nanocarbon Project, Nagoya University, Chikusa, Nagoya 464-8602, Japan

ABSTRACT

Atomically thin layered materials such as graphene and transition-metal dichalcogenides exhibit great potential as active materials in optoelectronic devices because of their high carrier-transporting properties and strong light–matter interactions. Here, we demonstrated that the photovoltaic performances of graphene/Si Schottky junction solar cells were significantly improved by inserting a chemical vapor deposition (CVD)-grown, large $MoS_2$ thin-film layer. This layer functions as an effective passivation and electron-blocking/hole-transporting layer. We also demonstrated that the photovoltaic properties are enhanced with increasing number of graphene layers and decreasing thickness of the $MoS_2$ layer. A high photovoltaic conversion efficiency of 11.1% was achieved with the optimized trilayer-graphene/$MoS_2$/$n$-Si solar cell.

TEXT

With an increase in low dimensional material research, these atomically thin materials have been widely and intensively studied from the viewpoints of both fundamental physics and applications.[1-8] Transition-metal dichalcogenides (TMDs) ($MX_2$; $M$ = Mo, W; $X$ = Se, S) are among the most attractive two-dimensional (2D) layered materials that can be thinned down to atomic-scale thickness.[9] Monolayer molybdenum disulfide ($MoS_2$), which is a typical and well-studied TMD system, is a direct bandgap semiconductor, whereas bulk $MoS_2$ is an indirect bandgap semiconductor. The direct-to-indirect energy-gap transition occurs in $MoS_2$ when it changes from a monolayer to a bilayer, and the optical bandgap changes from 1.8 eV in monolayer $MoS_2$ to 1.2 eV in bulk $MoS_2$.[3,10,11] The optical band gap associated with their energy structures in MoS$_2$ can be changed by manipulation of the layer numbers, which has potential applications of various optical and electronic devices.[12-14]

2D materials are used as building blocks to fabricate artificial multi-junction nanostructures because layered 2D materials with an atomically flat surface can be easily stacked by van der Waals interactions. Artificial multi-junction nanostructures composed of various 2D materials with novel characteristics can be fabricated. Indeed, graphene on *h*-BN exhibits an extremely high carrier mobility that exceeds 60,000 cm$^2$V$^{-1}$s$^{-1}$,[15] which is much larger than that of graphene on an SiO$_2$/Si substrate. Very high photo-responsivity has also been reported for optoelectronic devices based on the stacking of graphene and MoS$_2$ (graphene/MoS$_2$).[16-18] The graphene/MoS$_2$ heterostructure is considered a promising platform for photovoltaic applications.[19-22] The photovoltaic properties of solar cells based on low dimensional carbon materials and Si heterojunctions have been intensively studied.[23-26] In the case of graphene/Si solar cell, the approaches used to improve photovoltaic performance have primarily involved carrier doping in the graphene layer, the insertion of a oxide layer, or the deposition of an antireflection layer onto the surface of a solar cell.[27-33]

In this study, we investigate the photovoltaic properties of graphene/MoS$_2$/*n*-Si solar cells. The photovoltaic properties were considerably enhanced by the insertion of chemical vapor deposition (CVD)-grown large MoS$_2$ thin film, which functions as an effective passivation and electron-blocking/hole-transporting layer. The optimized graphene/MoS$_2$/*n*-Si solar cell exhibited a high photovoltaic conversion efficiency of 11.1%, which is a remarkable conversion efficiency in a photovoltaic device using MoS$_2$ thin film.

## RESULTS AND DISCUSSION

Figure 1(a) shows a schematic of the graphene/MoS$_2$/$n$-Si solar cell. It comprises the two types of 2D materials, CVD-grown MoS$_2$ film and graphene, stacked on a patterned $n$-type SiO$_2$/Si substrate (with a $\phi$ = 1 mm window). Large-scale MoS$_2$ films for solar cells were fabricated by the CVD method.[34-36] As a first step, an MoO$_3$ (99.98%, Sigma-Aldrich) film with the desired thickness was fabricated on an SiO$_2$/Si substrate using the thermal-evaporation technique under vacuum conditions (~10$^{-3}$ Pa); this MoO$_3$ film was subsequently sulfurized. The sulfurization step of the MoO$_3$ film was conducted using the CVD furnace. The SiO$_2$/Si substrate with an MoO$_3$ film was placed in the center of a quartz tube, and sulfur (>98%, Wako) was loaded at the upstream side. The substrate was heated by the CVD furnace to 750 °C at a heating rate of 15 °C/min and maintained at this temperature for 30 min. At the growth temperature (750 °C), sulfur was also heated to 190 °C using a ribbon heater, resulting in sulfur transport. All processes were performed under high-purity Ar gas flow flowing at 50 sccm under ambient conditions. We fabricated MoS$_2$ films with various thicknesses on SiO$_2$/Si substrates, as shown in Figure S1.

The graphene/MoS$_2$/$n$-Si solar cells were fabricated by transferring CVD-grown MoS$_2$ films and graphene films (single-layer graphene, Graphene Platform) onto the $n$-type Si substrate (1–5 $\Omega \cdot$cm$^{-1}$) with an ~300-nm-thick SiO$_2$ layer. A circular window with a diameter of 1 mm was patterned on the SiO$_2$/Si substrate. This Si substrate was dipped into buffered hydrofluoric acid (HF) solutions for several seconds to remove the natural oxidation layer on the window, which obstructs the electrical contact. The MoS$_2$ film was then transferred $via$ thermal release tape (REVALPHA; No. 3195MS, Nitto Denko). In this study, the transfer process using the thermal release tape was a more facile method compared with the conventional method of using a polymer support film.[37,38] In Figure S2, we show the MoS$_2$ transfer process in detail. For the fine transfer, the first key step is peeling the CVD-grown MoS$_2$ film from the growth substrate. The water droplet facilitates clear peeling because of the difference in hydrophobicities between the SiO$_2$/Si substrate and the MoS$_2$ film.[39] In this transfer method, the process is completed quickly, because the MoS$_2$ film can be released from the tape by heating, thereby eliminating the need to expose the film to a solution for a prolonged period.

Subsequently, we transferred graphene layers onto MoS$_2$/*n*-Si. Poly(bisphenol A carbonate) dissolved in chloroform (3 wt%) was spin-coated onto the single-layer graphene deposited onto copper foil for the transfer process. The graphene film was floated on FeCl$_3$ to etch out the copper foil and was subsequently transferred to the MoS$_2$/*n*-Si substrate.[37] The graphene/MoS$_2$/*n*-Si was baked at 120 °C for 5 min, and the polymer layer was then dissolved in chloroform. Finally, Au was deposited as an electrode onto the surface of the graphene/MoS$_2$/*n*-Si, with the exception of the window region, *via* the Ar-ion sputtering method. Indium (In), as the cathode, was soldered onto the back side of the solar cell. The process of fabricating the graphene/MoS$_2$/*n*-Si solar cell is described in the Supporting Information.

Figure 2(a) shows the Raman spectrum of the CVD-grown MoS$_2$ film. The spectrum shows several Raman peaks, including the in-plane $E^1_{2g}$ mode (~383 cm$^{-1}$) and the out-of-plane $A_{1g}$ (~408 cm$^{-1}$) mode of MoS$_2$. The frequency difference between the $E^1_{2g}$ and $A_{1g}$ Raman peaks (~25 cm$^{-1}$) are consistent with the previously reported values for bulk MoS$_2$,[40,41] which suggests that the MoS$_2$ film used in this study is relatively thick, as described below in detail. The Raman spectrum of the graphene/MoS$_2$ structures also indicates the vertical stacking of graphene/MoS$_2$ (see Supporting Figure S3).

Figure 2(b) shows the absorption spectrum of the MoS$_2$ film deposited on the quartz substrate. Distinct absorption peaks around 680 nm and 640 nm are observed in the spectrum, which are assigned as A exciton (680 nm) and B exciton (640 nm) peaks associated with the direct bandgap transitions at the K point of MoS$_2$ in the momentum space. Also, we observed the intense C peak at ~460 nm. This peak arises from the distinguishing band structure (band nesting), which is the parallel region of the conduction and valence bands.[20,42,43]

Figure 2(c) shows the surface morphology of MoS$_2$ films measured by atomic force microscopy (AFM). The surface of CVD-grown MoS$_2$ films shows the assembly of small grains with a typical lateral size of ~30 nm and a roughness of ~6.5 nm. We also investigated the thickness of the MoS$_2$ films using AFM. Figure S1(c) shows AFM images measured at the step edge of a typical MoS$_2$ film used for devices and deposited onto a Si substrate. The inset of Figure 2(d) shows the profile at the line in the figure. The thickness of this MoS$_2$ film was evaluated as ~17 nm on the basis of this height profile, which corresponds to a film with ~20 MoS$_2$ layers because the thickness of a monolayer of MoS$_2$ is ~0.7 nm.[40,44] The evaluated thickness is consistent with the Raman spectral data (Figure 2(a)) and with the absorption spectra (Figure 2(b)). In particular, we fabricated MoS$_2$ films with various thicknesses as shown in Figure S1(b) and (c).

We conducted X-ray photoelectron spectroscopy (XPS) of fabricated MoS$_2$ films to verify the absence of residual ingredients. The XPS spectrum in Figure 2(d) shows several peaks of the Mo 3d and S 2s core levels. The Mo 3d XPS spectrum is also shown in the inset of Figure 2(d), which indicates that the peaks at 162.6, 229.8, and 232.8 eV are associated with the S 2s, Mo 3d$_{5/2}$, and Mo 3d$_{3/2}$ core levels. The binding energies of these peaks are consistent with the previously reported values of Mo$^{4+}$ and S$^{2-}$ in MoS$_2$.[45,46] The Raman, optical absorption, AFM, and XPS data, as shown in Figure 2(a)–(d), certify that MoS$_2$ films were definitely deposited *via* the CVD method.

Figure 3(a) shows the current-density–voltage (*J–V*) curves of the monolayer-graphene/MoS$_2$/*n*-Si solar cell under dark and AM 1.5 illumination conditions. The *J–V* curve of the monolayer-graphene/MoS$_2$/*n*-Si solar cell under the dark condition (gray line) exhibits typical diode behavior. Furthermore, the monolayer-graphene/MoS$_2$/*n*-Si solar cell (yellow line) exhibits clear photovoltaic properties. The open-circuit voltage $V_{OC}$, short-circuit current density $J_{SC}$, and fill factor FF of the cell are 0.41 V, 13.1 mA·cm$^{-2}$, 0.25, respectively, which results in a photovoltaic conversion efficiency ($\eta$) of 1.35%. The inset of Figure 3(a) shows the *J–V* curves of a graphene/*n*-Si solar cell without an MoS$_2$ layer, where the fabrication process and materials were same as those used to fabricate the monolayer-graphene/MoS$_2$/*n*-Si cell. The photovoltaic parameters, $V_{OC}$, $J_{SC}$, and FF are 0.22 V, 1.9 mA·cm$^{-2}$, and 0.18, respectively, which result in a photovoltaic conversion efficiency ($\eta$) of 0.07%. All the photovoltaic parameters of the graphene/MoS$_2$/*n*-Si solar cell are increased in comparison to those of the graphene/*n*-Si solar cell in our experiments, which suggests that the MoS$_2$ layer plays an important role in enhancing the photovoltaic performance of solar cells.

We varied the layer thicknesses of MoS$_2$ in the graphene/MoS$_2$/*n*-Si solar cells. Figure 3 shows the *J–V* curves of graphene/MoS$_2$/*n*-Si solar cells with different numbers of graphene layers when illuminated under the AM 1.5 condition. The red, orange, and yellow lines in Figure 3(b) corresponds to *J–V* curve of graphene/MoS$_2$/*n*-Si solar cells with monolayer, bilayer, and trilayer graphene, respectively, where we controlled the number of graphene layers by changing the number of times the graphene transfer process was conducted. The photovoltaic properties of

graphene/MoS$_2$/*n*-Si solar cells increased with increasing number of graphene layers. The trilayer-graphene/MoS$_2$/*n*-Si solar cells exhibit a $V_{OC}$, $J_{SC}$, and FF of 0.54 V, 28.1 mA·cm$^{-2}$, and 0.53, which results in a high photovoltaic efficiency of 8.0%.

These parameters of monolayer- and bilayer-graphene/MoS$_2$/*n*-Si solar cells and the series resistance estimated from the slope of each *J–V* curve are shown in Table 1 of Supporting Information S4. The series resistance of the solar cells drastically decreased with increasing number of graphene layers, consistent with the previously reported results of decreased sheet resistance in graphene layers.[47] Our results suggest that the decrease in series resistance is responsible for the reduction of loss, which contributes to the enhancement of photovoltaic conversion efficiency in graphene/MoS$_2$/*n*-Si solar cells.

The incident photon to current conversion efficiency (IPCE) spectra revealed that MoS$_2$ functions as a shading layer at the point of light absorption, as shown in Supporting Information S5. Thus, the generated carriers in the *n*-Si layer are the main contributors to the photovoltaic current in the cell. Therefore, the enhancement of photovoltaic performance by the insertion of an MoS$_2$ layer implies that the MoS$_2$ layer serves positive roles as both an effective passivation layer and a hole-transporting/electron-blocking layer in the photovoltaic process, as described below in detail.

The MoS$_2$ plays an important role of passivation layer for separation the graphene and Si to reduce the interface carrier recombination. The tunneling barrier effect by the passivation layer with a thickness of several nanometers to several tenths of nanometers prevents direct contact between the semiconductor and conductive layers and can thereby suppress the recombination losses at the interface. The tunneling barrier thickness also plays a key role in determining device performance.[48] We fabricated a trilayer-graphene/MoS$_2$/n-Si solar cell using a thinner (~9 nm) MoS$_2$ film. Figure 4 shows the *J–V* curve for the graphene/MoS$_2$/n-Si solar cell. The optimized graphene/MoS$_2$/n-Si solar cells exhibit $V_{OC}$, $J_{SC}$, and FF photovoltaic parameters of 0.56 V, 33.4 mA·cm$^{-2}$, and 0.60, which results in a high photovoltaic efficiency of 11.1%. This value of 11.1% is among the highest efficiencies reported for a solar cell containing an MoS$_2$ layer. Thinner layer is suitable not only for passivation effect but also for shadow effect by MoS$_2$ layer. This improvement indicates that more photons passed through the MoS$_2$ and were absorbed by the Si.

To better understand the photovoltaic process, we constructed band alignment diagrams for the graphene/*n*-Si and graphene/MoS$_2$/*n*-Si solar cells, as shown in Figure 5. Figure 5(a) shows the band alignment of a graphene/*n*-Si solar cell under ambient conditions, where the graphene has been *p*-doped *via* the FeCl$_3$ etching process.[49-51] The photovoltaic process in the graphene/*n*-Si solar cells is understood by the Schottky barrier formed at the semimetal (graphene) and semiconductor (*n*-Si) interface. Figure 5(b) shows a schematic band diagram for graphene/MoS$_2$/*n*-Si solar cells. We confirmed that the CVD-grown MoS$_2$ exhibits *n*-type properties on the basis of the transfer characteristics of the electric double-layer transistor (Supporting Information S6).[52] The insertion of an MoS$_2$ layer into graphene/*n*-Si solar cells modifies the band alignment. Indeed, the photovoltaic properties are enhanced by the insertion of an MoS$_2$ layer, as shown in Figure 3(a). The total built-in voltage ($V_{bi}$) of graphene/MoS$_2$/*n*-Si solar cells is greater than that of the graphene/n-Si solar cell because the Schottky barrier height is determined by the energy difference between the electron affinity $\chi_s$ of the semiconductor and the work function $W_m$ of graphene, *i.e.*, $W_m - \chi_s$.[53-55] The increase of the built-in field by insertion of an MoS$_2$ layer can directly improve the $V_{OC}$ and the photovoltaic conversion efficiency. Moreover, because of the difference between Fermi levels, the bottom of the conduction band and the top of the valence band of the $MoS_2$ layer at the interface of graphene are shifted upward, resulting in the energy barrier at the interface between $MoS_2$ and $n$-Si. The energy barrier of the $MoS_2$ layer functions as an effective electron blocking layer for the photogenerated electrons in the Si layer. Furthermore, in the case of the photogenerated hole, the energy barrier created by the $MoS_2$ layer functions as an effective hole-transport layer. The effective carrier transporting (blocking) layer of $MoS_2$ effectively reduces the recombination loss of carriers, which in turn results in an increase of the $J_{SC}$. Therefore, the $MoS_2$ layer between the graphene and $n$-Si contribute to the enhancement of photovoltaic performance as a carrier transporting (blocking) layer, as shown in Figure 3(a).

CONCLUSIONS

In summary, we investigated the considerable enhancement of the photovoltaic performance of graphene/Si solar cells using the 2D material $MoS_2$. The large-area CVD-grown $MoS_2$ films enabled the fabrication of graphene/$MoS_2$/$n$-Si solar cells. The $MoS_2$ acts as both an effective passivation layer and hole-transporting/electron-blocking layer, which contributes to the remarkable conversion efficiency of 11.1% in the fabricated graphene/$MoS_2$/$n$-Si solar cells. This result opens new directions for $MoS_2$ research and the potential use of $MoS_2$ in various optoelectronic applications.

MATERIALS AND METHODS

Raman spectrum of the MoS$_2$ films was obtained using a laser Raman microscope (Nanophoton, RAMANtouch). The optical absorption spectrum of the MoS$_2$ films on the glass substrates was measured using a UV/Vis spectrophotometer (SHIMADZU, UV-1800). XPS was conducted on a TRXPS spectrometer (JEOL, JPS-9010TRX). The surface morphologies were measured by AFM (Agilent Technologies). To test the photovoltaic performance, the solar cells were irradiated using a solar simulator (San-Ei Electric XES-40S1) under AM 1.5 conditions (100 mW·cm$^{-2}$) and the current density–voltage data were recorded using a source meter (Keithley 2400). The AM 1.5 condition for the solar simulator was confirmed using a standard cell (BS-500BK). For the incident-to-photon conversion efficiency (IPCE) measurements, the devices were tested using a monochromated xenon arc light system (Zolix LSP-X150).

*Conflict of Interest*: The authors declare no competing financial interest.

FIGURES

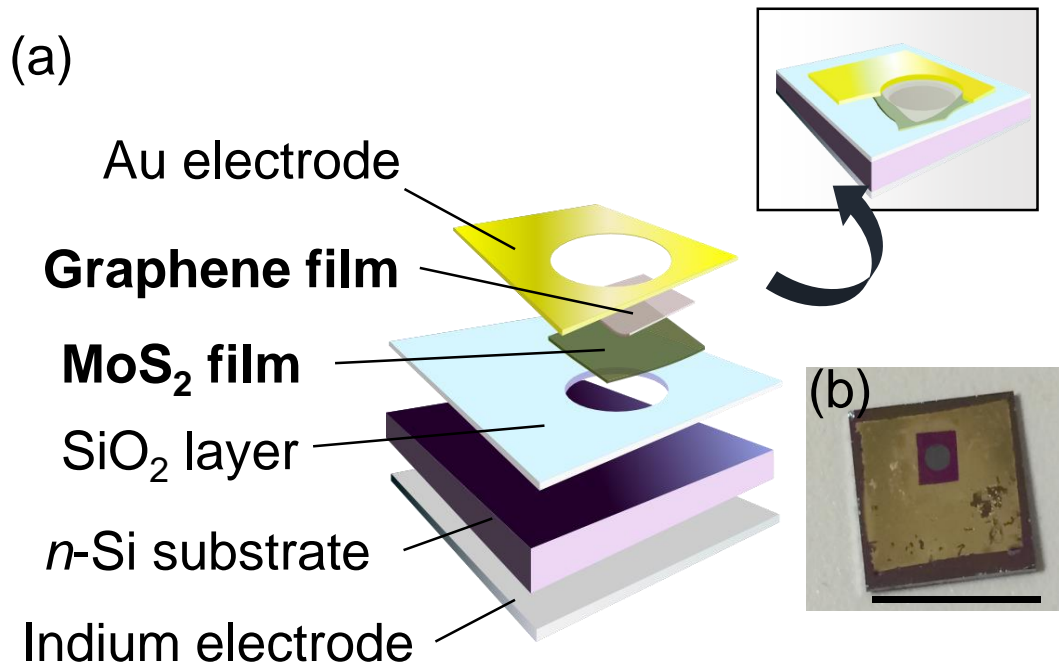

Fig. 1 Y. Tsuboi *et al*.

**Figure 1**. (a) Schematic of the graphene/MoS$_2$/*n*-Si solar cell. (b) An optical image of a graphene/MoS$_2$ thin film deposited onto a substrate. The scale bar is 1 cm.

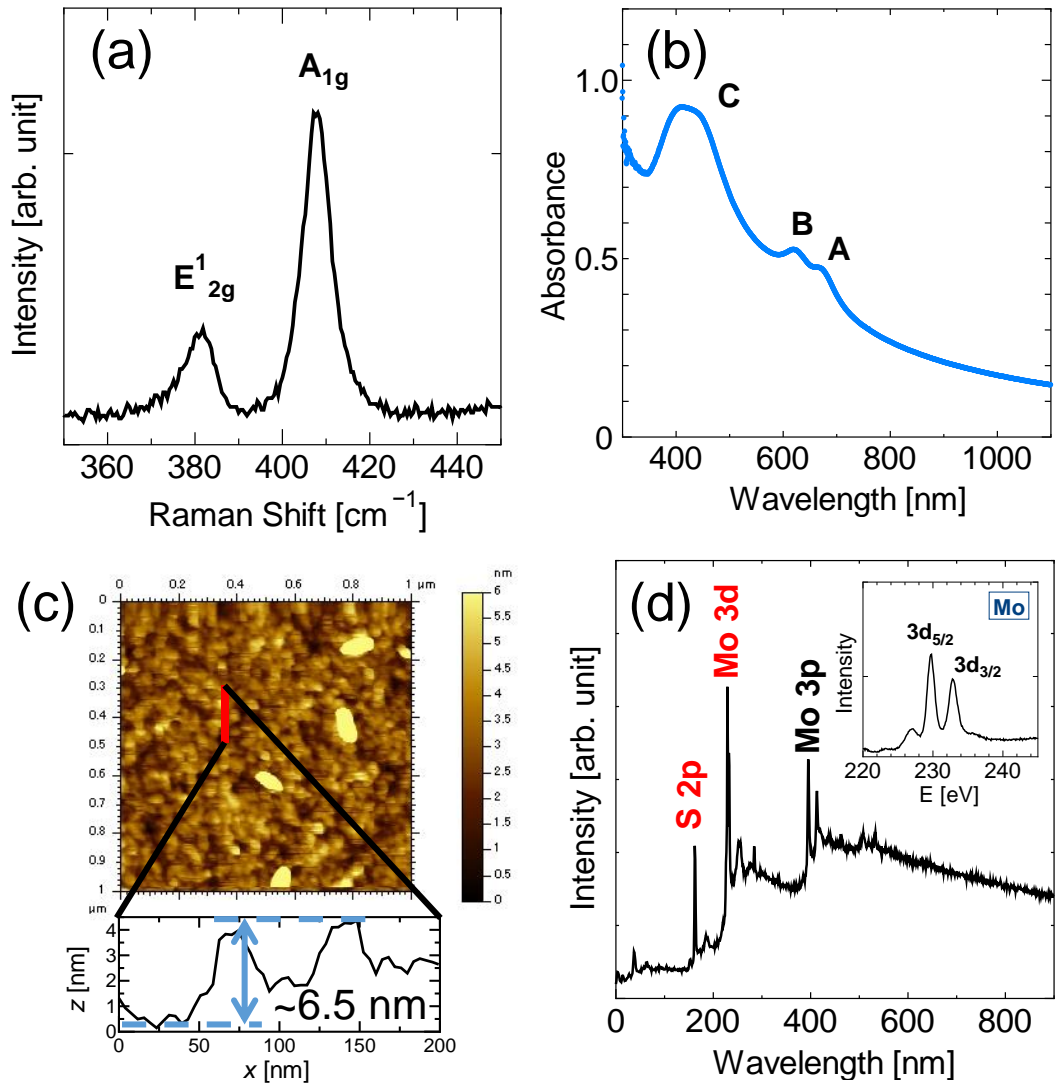

Fig. 2 Y. Tsuboi *et al.*

**Figure 2.** Characterization of CVD-grown $MoS_2$ film. (a) Raman spectra of $MoS_2$ films. The two Raman peaks associated with $MoS_2$, an in-plane ($E^1_{2g}$) mode at 383 cm$^{-1}$ and an out-of-plane ($A_{1g}$) mode at 408 cm$^{-1}$, are observed. (b) Optical absorption spectra of $MoS_2$. Distinct absorption peaks associated with A and B excitons are observed. (c) AFM image of an as-grown $MoS_2$ film. The height profile in the line is shown in the Figure. (d) X-ray photoelectron spectrum of an $MoS_2$ film showing Mo and S peaks. The inset is the Mo 3d spectrum.

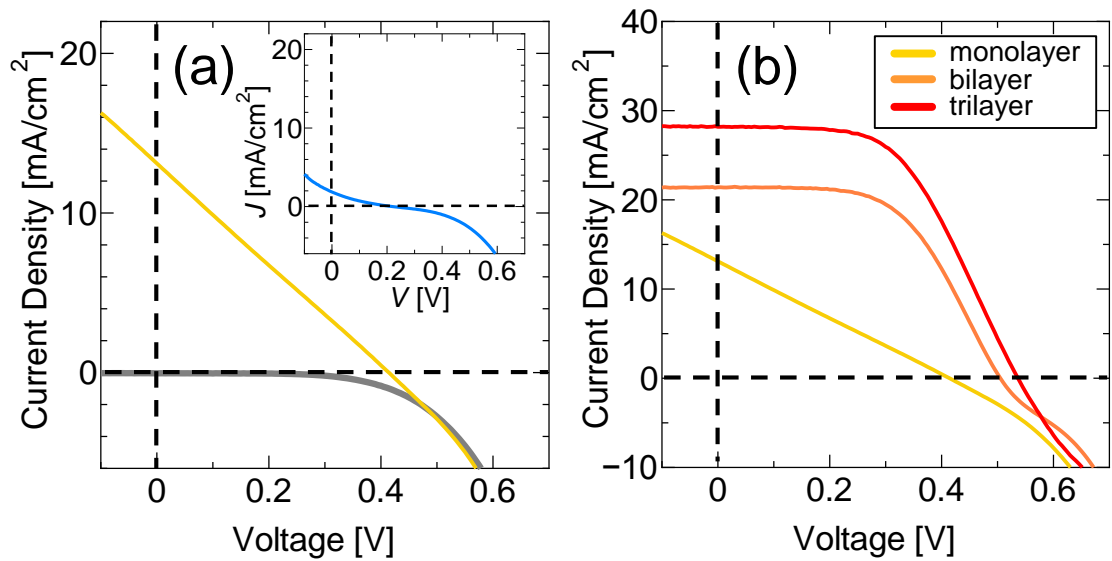

Fig. 3 Y. Tsuboi *et al.*

**Figure 3.** Photovoltaic performance of monolayer-, bilayer-, and trilayer-graphene/$MoS_2$/$n$-Si solar cells. (a) Current density–voltage ($J$–$V$) curves of a monolayer-graphene/$MoS_2$/$n$-Si solar cell under darkness (gray line) and under AM 1.5 illumination conditions (yellow line). The $J$–$V$ curve of a bilayer-graphene/$n$-Si solar cell under AM 1.5 illumination is also shown in the inset. (b) $J$–$V$ curves of monolayer-, bilayer-, and trilayer-graphene/$MoS_2$/$n$-Si solar cells, where the number of graphene layers was controlled by the number of graphene transfer processes performed.

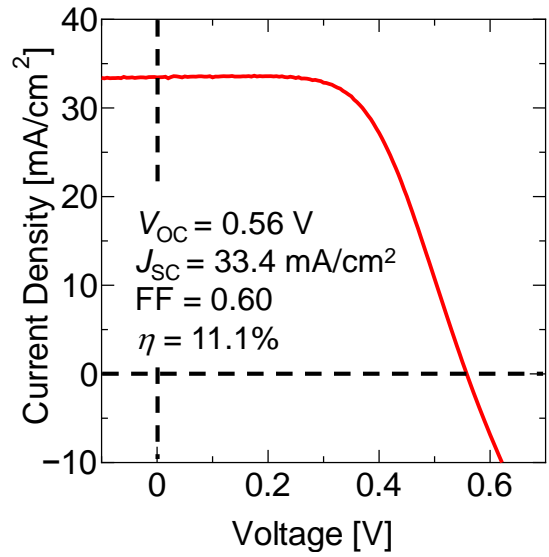

Fig. 4  Y. Tsuboi *et al.*

**Figure 4.** The current density–voltage (*J–V*) curves of a trilayer-graphene/$MoS_2$/*n*-Si solar cell with a 9-nm-thick $MoS_2$ layer; the curves were measured under AM 1.5 illumination conditions. Photovoltaic parameters, including the open-circuit voltage $V_{OC}$, short-circuit current density $J_{SC}$, fill factor FF, and photovoltaic efficiency $\eta$ determined from this curve are indicated in the Figure.

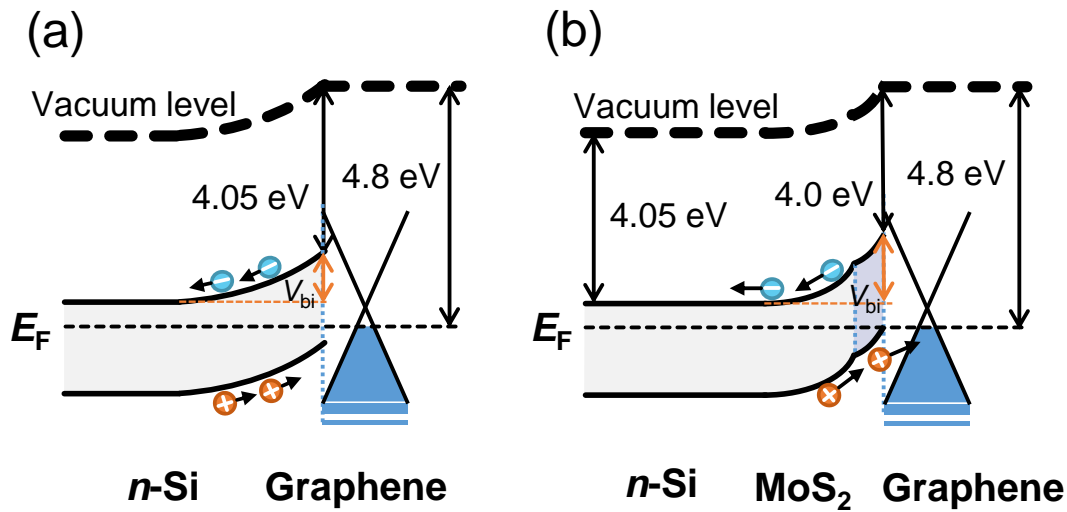

Fig. 5 Y. Tsuboi *et al.*

**Figure 5.** Schematics of band diagrams for the solar cells. The photovoltaic processes (a) in the graphene/*n*-Si and (b) in the graphene/MoS$_2$/*n*-Si solar cells are shown.

AUTHOR INFORMATION

**Corresponding Author**

matsuda@iae.kyoto-u.ac.jp

ACKNOWLEDGMENT

Part of this work was supported by M. Endo, A. Wakamiya, Y. Murata, T. Nakata, and T. Morii for experimental equipment. We also thank to Nitto Denko Corporation for providing us thermal release tape. This study was supported by a Grant-in-Aid for Scientific Research from the Japan Society for Promotion of Science (Grants Nos. 24681031, 22740195, 22016007, 25000003, and 23340085), the Precursory Research for Embryonic Science and Technology program from the Japan Science and Technology Agency, the Yazaki Memorial Foundation for Science and Technology, the Nippon Sheet Glass Foundation for Materials Science and Engineering, and The Canon Foundation. KF acknowledges the Leading Graduate Program in Science and Engineering, Waseda University from MEXT, Japan.*Supporting Information Available*: Characterization of $MoS_2$ film, process-flow of $MoS_2$ film transfer, photovoltaic parameters of graphene/$MoS_2$/*n*-Si solar cells, and carrier transport properties of $MoS_2$ films.

17
23340085), the Precursory Research for Embryonic Science and Technology program from the Japan Science and Technology Agency, the Yazaki Memorial Foundation for Science and Technology, the Nippon Sheet Glass Foundation for Materials Science and Engineering, and The Canon Foundation. KF acknowledges the Leading Graduate Program in Science and Engineering, Waseda University from MEXT, Japan.

*Supporting Information Available*: Characterization of $MoS_2$ film, process-flow of $MoS_2$ film transfer, photovoltaic parameters of graphene/$MoS_2$/*n*-Si solar cells, and carrier transport properties of $MoS_2$ films.

REFERENCES

1.      Geim, A. K.; Novoselov, K. S., The Rise of Graphene. *Nat. Mater.* **2007**, 6, 183-191.

2.      Splendiani, A.; Sun, L.; Zhang, Y.; Li, T.; Kim, J.; Chim, C.-Y.; Galli, G.; Wang, F., Emerging Photoluminescence in Monolayer $MoS_2$. *Nano Lett.* **2010**, 10, 1271-1275.

3.      Radisavljevic, B.; Radenovic, A.; Brivio, J.; Giacometti, V.; Kis, A., Single-layer $MoS_2$ Transistors. *Nat. Nanotechnol.* **2011**, 6, 147-150.

4.      Rao, C. N. R.; Sood, A. K.; Subrahmanyam, K. S.; Govindaraj, A., Graphen, Das Neue Zweidimensionale Nanomaterial. *Angew. Chem.* **2009**, 121, 7890-7916.

# Supporting Information

## Enhanced Photovoltaic Performances of Graphene/Si Solar Cells by Insertion of an MoS$_2$ Thin Film

Yuka Tsuboi[†], Feijiu Wang[†], Daichi Kozawa[†], Kazuma Funahashi[‡],

Shinichiro Mouri[†], Yuhei Miyauchi[†], Taishi Takenobu[‡] and Kazunari Matsuda[†]*

[†]*Institute of Advanced Energy, Kyoto University, Gokasho, Uji, Kyoto 611-0011, Japan*

[‡]*Department of Advanced Science and Engineering, Waseda University, Shinjuku-ku, Tokyo 169-8555, Japan*

[§]*JST, ERATO, Itami Molecular Nanocarbon Project, Nagoya University, Chikusa, Nagoya 464-8602, Japan*

*Email: matsuda@iae.kyoto-u.ac.jp

## S1. MoS$_2$ film fabrication

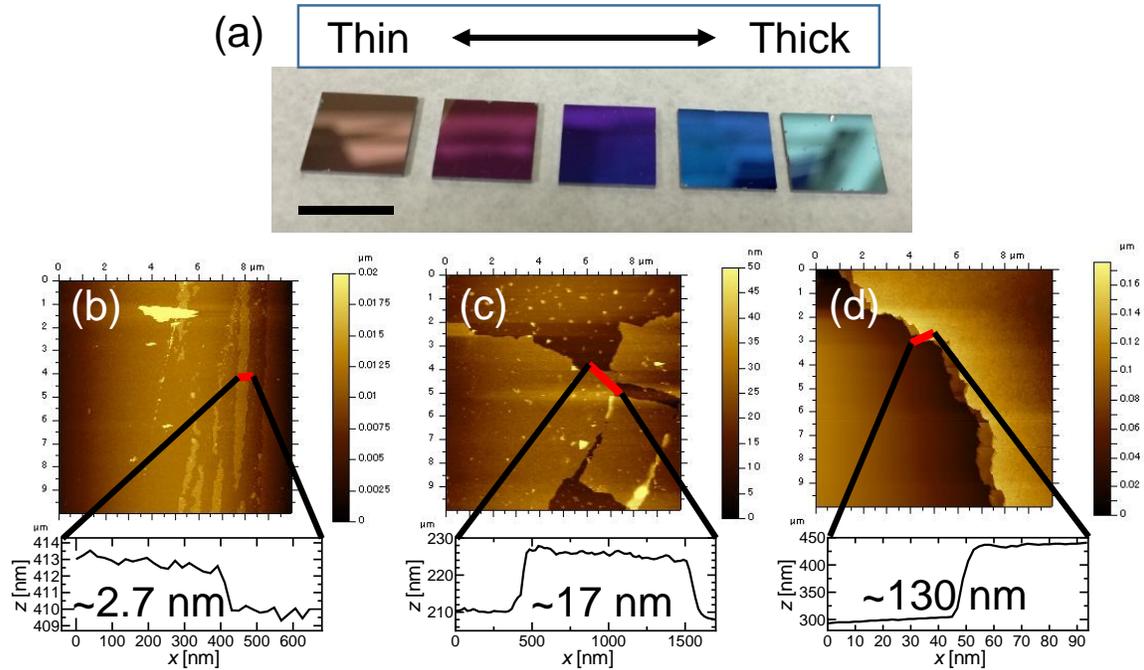

**Figure S1.** (a) Optical image of as-grown MoS$_2$ films with various layer thicknesses on SiO$_2$/Si substrates, as fabricated by the CVD method. The scale bar is 1 cm. The thickness of the MoS$_2$ films was controlled *via* the MoO$_3$ deposition time. The leftmost image is an SiO$_2$/Si substrate. (b)–(d) AFM images of an MoS$_2$ film. The height profiles are measured at the step edge between the MoS$_2$ film and the SiO$_2$/Si substrate. The thickness of the MoS$_2$ films was varied from 2.7 to 130 nm. For solar cells in Figure 3, we used films with ~17-nm thickness, as shown in (c).

## S2. Process-flow of MoS$_2$ film transfer

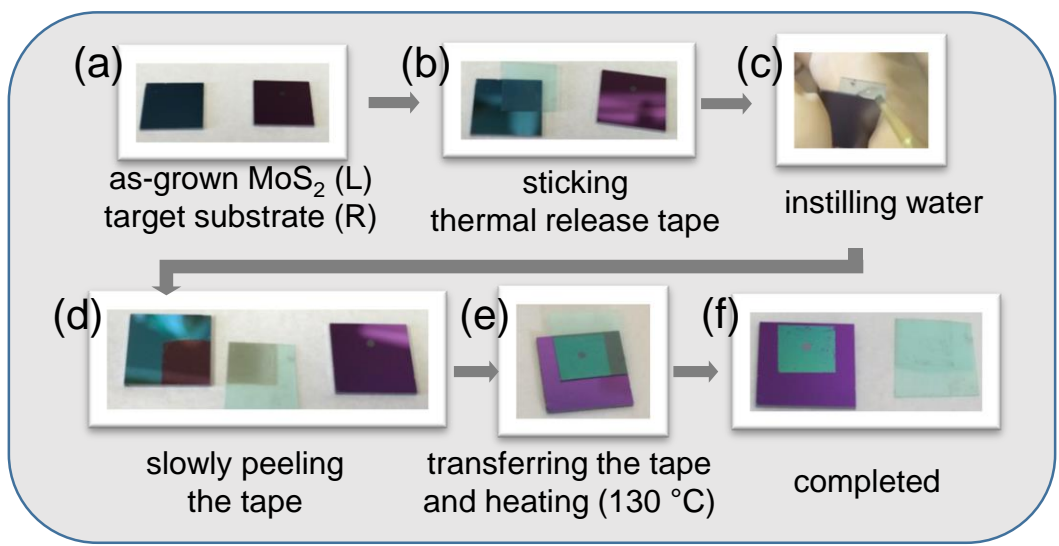

**Figure S2.** (a)–(f) The process-flow of as-grown MoS$_2$ film transfer. This method was conducted using thermal release tape (REVALPHA; No. 3195MS, Nitto Denko). The sticking MoS$_2$ film with the thermal release tape (left in (b)) was easily released by a drop of water, as shown in (c). After the tape with MoS$_2$ was transferred to the target substrate, the tape was removed by heating the substrate on a hotplate.

## S3. Characterization of stacked graphene/MoS$_2$ film

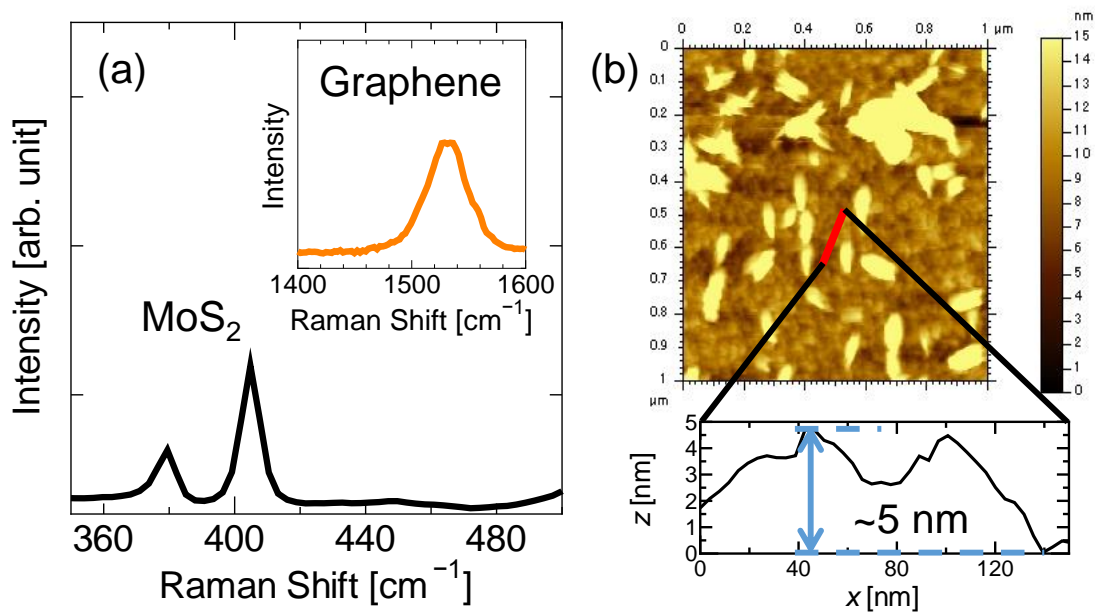

**Figure S3.** (a) Raman spectrum of the vertical stacking of graphene/MoS$_2$. (b) AFM image of a stacked graphene/MoS$_2$ film and its height profile along the red line in the image. The surface roughness of the graphene/MoS$_2$ film decreases in comparison with that of the MoS$_2$ film (Figure 2(a)).

## S4. Photovoltaic parameters of graphene/MoS$_2$/$n$-Si solar cells

**Table S4.** Photovoltaic parameters of graphene/MoS$_2$/$n$-Si solar cells with monolayer-graphene (MLG), bilayer-graphene (BLG) and trilayer-graphene (TLG), as evaluated from experimentally obtained $J$–$V$ curves. The open-circuit voltage $V_{OC}$, short-circuit current density $J_{SC}$, fill factor FF, photovoltaic efficiency $\eta$, and series resistance $R_S$ are shown. Here, the $R_S$ values were estimated from the slopes of the $J$–$V$ curves.[S1]

| Cell Type | $V_{OC}$ [V] | $J_{SC}$ [mA/cm$^2$] | FF | $\eta$ [%] | $R_s$ [W·cm$^2$] |
|---|---|---|---|---|---|
| MLG/MoS$_2$/$n$-Si | 0.41 | 13.1 | 0.25 | 1.35 | 30 |
| BLG/MoS$_2$/$n$-Si | 0.51 | 21.4 | 0.55 | 5.98 | 11 |
| TLG/MoS$_2$/$n$-Si | 0.54 | 28.2 | 0.53 | 8.02 | 8.8 |

## S5. The incident photon-to-current conversion efficiency (IPCE) spectrum of graphene/MoS$_2$/*n*-Si solar cells

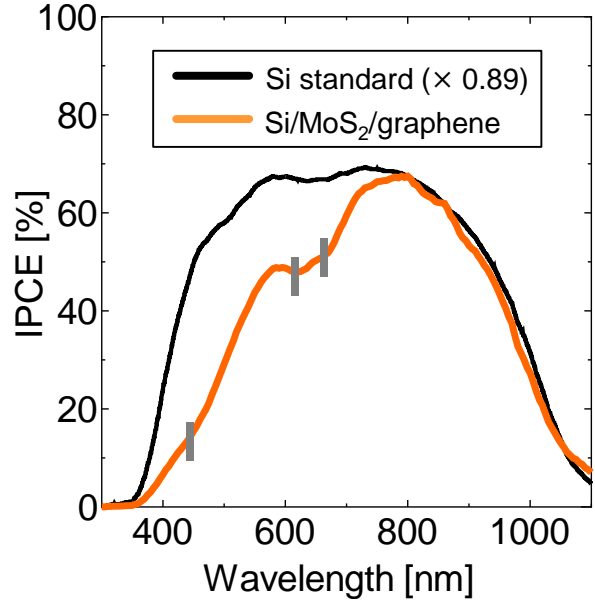

**Figure S5.** IPCE spectra of a trilayer-graphene/MoS$_2$/*n*-Si solar cell and a standard Si solar cell. The spectrum of the standard Si solar cell is normalized by the values at 800 nm. The dip structures were observed in the IPCE spectrum of graphene/MoS$_2$/*n*-Si in the shorter-wavelength region (400–700 nm). The positions of these dip structures are consistent with the absorption peaks of the MoS$_2$ film shown in Figure 2(b), which suggests that the generated carriers in the *n*-Si layer are the primary contributors to the photovoltaic current in the cell, whereas the MoS$_2$ functions as a shading layer at the point of light absorption.

## S6. Carrier transport properties of an $MoS_2$ film, as measured by an electric double-layer transistor

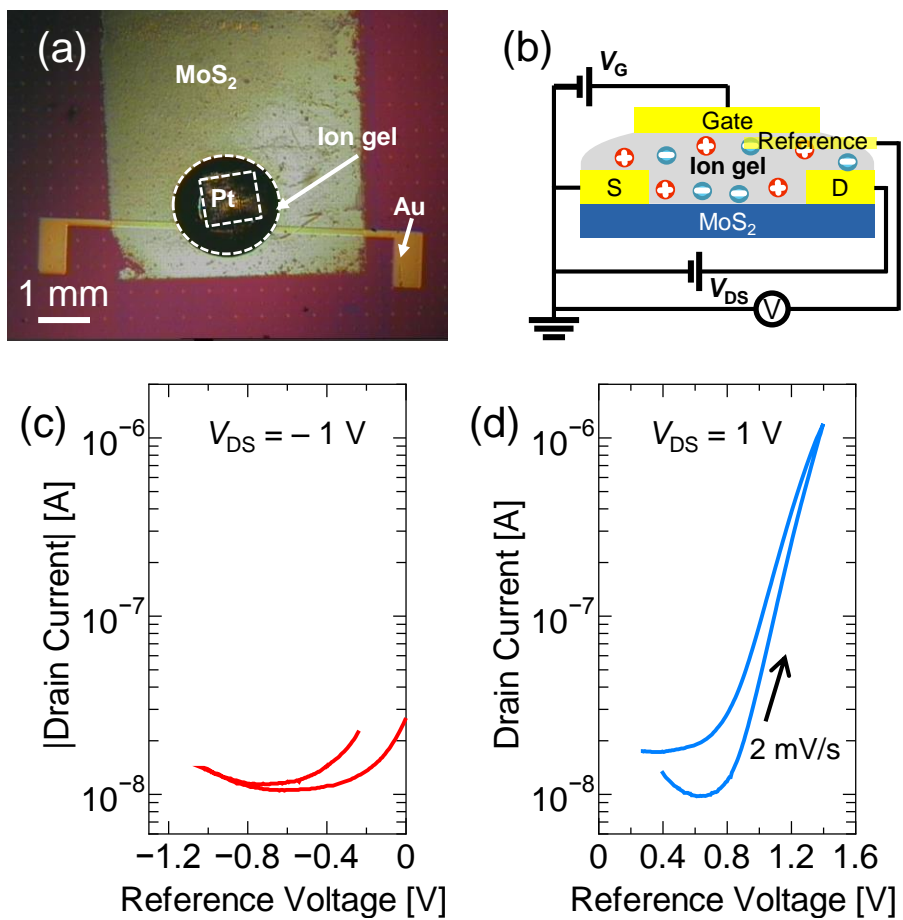

**Figure S6.** (a) Optical image of CVD-grown $MoS_2$ electric double-layer transistor and (b) A schematic illustration of the transistor, in which an ionic gel ([EMIM][TFSI] with PS-PMMA-PS) as a gate electrolyte, Pt as a top-gate electrode, and Ni/Au source and drain electrodes were used.[S2] (c)(d) Transfer characteristics measured by changing the voltage of the Pt top-gate electrodes. Current resulting from electron transfer was confirmed (blue line), whereas current resulting from hole transfer was not (red line).